\documentclass{iaus}
\usepackage{graphicx}
\newcommand{\Al}{$^{26\!}$Al\ }
\newcommand{\Fe}{$^{60\!}$Fe\ }

\title[$^{60}$Fe and Massive Stars] 
{$^{60}$Fe and Massive Stars}

\author[W. Wang]   
{Wei Wang}

\affiliation{National Astronomical Observatories, Chinese Academy
of Sciences, Beijing 100012, China \break email:
wangwei@bao.ac.cn}

\pubyear{2008}
\volume{252}  
\pagerange{}
\date{}
\jname{Proceedings Title IAU Symposium}
\editors{Licai Deng, K.L. Chan, C. Chiosi , eds.}
\begin{document}

\maketitle

\begin{abstract}
Gamma-ray line emission from radioactive decay of $^{60}$Fe
provides constraints on nucleosynthesis in massive stars and
supernovae. We detect the $\gamma$-ray lines from $^{60}$Fe decay
at 1173 and 1333~keV using three years of data from the
spectrometer SPI on board {\em INTEGRAL\/}. The average flux per
line is $ (4.4 \pm 0.9) \times 10^{-5} ~{\rm ph} ~\hbox{cm}^{-2}
\hbox{s}^{-1}~\hbox{rad}^{-1}$ for the inner Galaxy region.
Deriving the Galactic $^{26}$Al gamma-ray line flux with using the
same set of observations and analysis method, we determine the
flux ratio of $^{60}{\rm Fe}/^{26}{\rm Al}$ gamma-rays as $0.15
\pm 0.05$. We discuss the implications of these results for the
widely-held hypothesis that $^{60}$Fe is synthesized in
core-collapse supernovae, and also for the closely-related
question of the precise origin of $^{26}$Al in massive stars.

\end{abstract}

\firstsection 
\section{Introduction}

The radioactive isotope $^{60}$Fe is believed to be synthesized
through successive neutron captures on Fe isotopes (e.g.,
$^{56}$Fe) in a neutron-rich environment inside He burning shells
in AGB stars (\Fe is stored in white dwarfs and cannot be ejected)
and massive stars, before or during their final evolution to core
collapse supernovae (CCSN). \Fe can be also synthesized in Type Ia
SNe (Woosley 1997). It is also destroyed by the \Fe($n,\gamma$)
process. Since its closest parent, $^{59}$Fe is unstable, the
$^{59}$Fe($n,\gamma$) process must compete with the
$^{59}$Fe($\gamma^-$) decay to produce an appreciate amount of
\Fe.

The decay chains of \Fe are shown in Figure 1. \Fe firstly decays
to $^{60}$Co, with emitting $\gamma$-ray photons at 59 keV, and
then decays to $^{60}$Ni, with emitting $\gamma$-ray photons at
1173 and 1333 keV. The gamma-ray efficiency of the 59 keV
transition is only $\sim 2\%$ of those at 1173 and 1333 keV, so
the gamma-ray flux at 59 keV is much lower than the fluxes of the
high energy lines. The 59 keV gamma-ray line is very difficult to
be detected with present missions. Measurements of the two high
energy lines have been the main scientific target to study the
radioactive \Fe isotope in the Galaxy.

\Fe has been found to be part of meteorites formed in the early
solar system (Shukolyukov et al. 1993). The inferred \Fe/$^{56}$Fe
ratio for these meteorites exceeded the interstellar-medium
estimates from nucleosynthesis models, which led to suggestion
that the late supernova ejection of \Fe occurred before formation
of the solar system (Tachibana et al. 2006). Yet, this is a proof
for cosmic \Fe production, accelerator-mass spectroscopy of
seafloor crust material from the southern Pacific ocean has
revealed an \Fe excess in a crust depth corresponding to an age of
2.8 Myr (Knie et al. 2004). From this interesting measurement, it
is concluded that a supernova explosion event near the solar
system occurred about 3 Myr ago, depositing some of its debris
directly in the earth's atmosphere. All these measurements based
on material samples demonstrate that \Fe necleosynthesis does
occur in nature. It is now interesting to search for current \Fe
production in the Galaxy through detecting radioactive-decay
$\gamma$-ray lines.

\begin{figure}
\begin{center}
 \includegraphics[width=10cm]{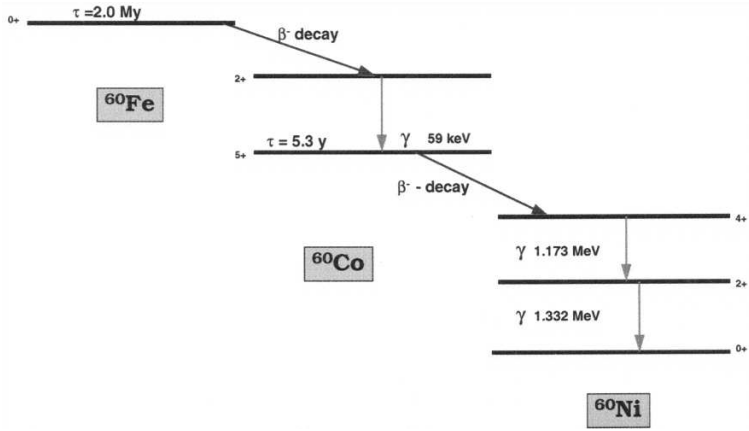}
  \caption{The decay scheme of \Fe. The mean lifetime is $2\times
10^6$ years. The gamma-ray flux at 59 kev line is $\sim 2\%$ of
those at 1173 and 1333 keV.}
\end{center}
\end{figure}

\section{\Fe emission in the Galaxy}

Due to its long decay time ($\tau \simeq$2.2~My), $^{60}$Fe
survives to be detected after the supernova ejected it into the
interstellar medium, by $\beta$-decay via $^{60}$Co and $\gamma$
emission at $1173$ keV and 1333 keV -- like other radioactive
isotopes: $^{44}$Ti, $^{56,57}$Co, and $^{26}$Al. These isotopes
provide evidence that nucleosynthesis is ongoing in the Galaxy
(The et al. 2006; Diehl et al. 2006). Specially, measurements of
$^{60}$Fe promise to provide new information about the massive
star nucleosynthesis in the late pre-supernova stages.

Gamma-ray signal of \Fe from the sky is very weak, so there are no
confident detections of \Fe in the Galaxy reported in the previous
measurements. Recently, RHESSI reported observations of the
gamma-ray lines from $^{60}$Fe with an average flux of $ (6.3\pm
5.0) \times 10^{-5} {\rm ph\ cm^{-2}\ s^{-1}}$ (Smith 2004).

Now, the spectrometer aboard INTEGRAL (SPI) operates on space. The
{\em INTE}rnational {\em G}amma-{\em R}ay {\em A}strophysics {\em
L}aboratory (INTEGRAL) is an European (ESA) Gamma-Ray Observatory
Satellite Mission for the study of cosmic gamma-ray sources in the
keV to MeV energy range (Winkler et al. 2003). INTEGRAL was
successfully launched from Baikonur Cosmodrome (Kazakhstan) on
October 17, 2002. The INTEGRAL orbit is eccentric, with an apogee
of 153 000 km, a perigee of 9000 km, and a 3 day period. INTEGRAL
will continue to work until 2012 approved by ESA. SPI/INTEGRAL
consists of 19 high purity germanium detectors which allow for
high spectral resolution of $\sim 2.5$ keV at 1 MeV, suitable for
astrophysical studies of individual gamma-ray lines and their
shapes, e.g. the 511 keV line, $\gamma$-ray lines from
radioactivities of $^{44}$Ti, \Al and \Fe. The basic measurement
of SPI consists of event messages per photon triggering the Ge
detector camera. We distinguish events which trigger a single Ge
detector element only ({\it single event}, SE), and events which
trigger more than two Ge detector elements nearly simultaneously
({\it multiple event}, ME).

We analyzed the first year of INTEGRAL data to detect the
$\gamma$-ray lines from $^{60}$Fe with an average line flux of
$(3.7 \pm\ 1.1) \times 10^{-5} {\rm ph\ cm^{-2}\ s^{-1}}$ (Harris
et al. 2005). But the strong background lines near $^{60}$Fe lines
still contaminate the spectra, which makes this preliminary
results questionable. At present, we use three years of INTEGRAL
data (from 2003.3 - 2006.3), aiming at a consolidation of the
INTEGRAL/SPI measurement of $^{60}$Fe gamma-rays.

The newest results on $^{60}$Fe gamma-ray lines by INTEGRAL/SPI
are presented in Figure 2. All the fluxes given by different
databases are consistent with each other. The strong background
line at 1337 keV has also been eliminated rather well.
Furthermore, a superposition of the four spectra of Figure 2 is
shown in Figure 3. The line flux estimated from the combined
spectrum is $(4.4\pm 0.9)\times 10^{-5}\mathrm{ ph\ cm^{-2}\
s^{-1}\ rad^{-1}}$. Our significance estimate for the combined
spectrum is $\sim 5\sigma$ (Wang et al. 2007).

\begin{figure}
 \includegraphics[width=14cm]{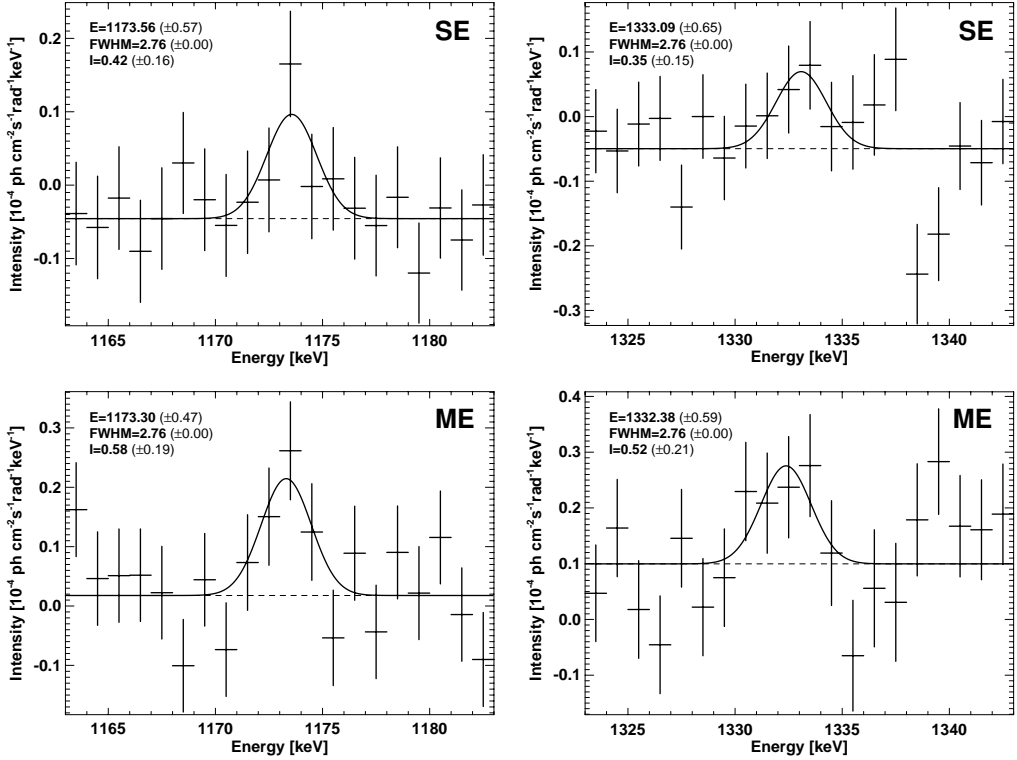}
  \caption{The spectra of two gamma-ray lines of $^{60}$Fe from the
inner Galaxy: 1173 keV and 1333 keV. We have shown the results
both from SE and ME databases. For the SE database, we find a line
flux of $(4.2\pm 1.6)\times 10^{-5}{\rm ph\ cm^{-2}\ s^{-1}\
rad^{-1}}$ for the 1173 keV line and $ (3.5\pm 1.5)\times
10^{-5}{\rm ph\ cm^{-2}\ s^{-1}\ rad^{-1}}$ for the 1333 keV line.
For the ME database, the line flux is $ (5.8\pm 1.9)\times
10^{-5}{\rm ph\ cm^{-2}\ s^{-1}\ rad^{-1}}$ for the 1173 keV line
and $ (5.2\pm 2.1)\times 10^{-5}{\rm ph\ cm^{-2}\ s^{-1}\
rad^{-1}}$ for the 1333 keV line }
\end{figure}

\begin{figure}
\centering
\includegraphics[width=9cm]{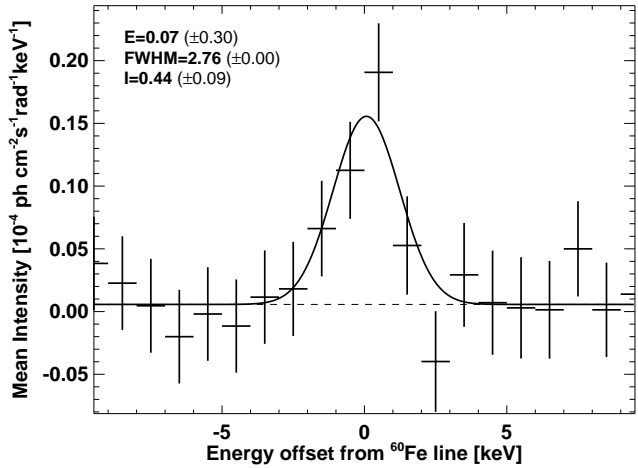}
\caption{The combined spectrum of the $^{60}$Fe signal in the
inner Galaxy, superimposing the four spectra of Figure 2. In the
laboratory, the line energies are 1173.23 and 1332.49 keV; here
superimposed bins are zero at 1173 and 1333 keV. We find a
detection significance of 5$\sigma$. The average line flux is
estimated as $ (4.4\pm 0.9)\times 10^{-5}\mathrm{ ph\ cm^{-2}\
s^{-1}\ rad^{-1}}$. }
\end{figure}

\section{The ratio of \Fe/\Al}

\Al is an unstable isotope with a mean lifetime of 1.04 Myr. \Al
can first decay into an excited state of $^{26}$Mg, which
de-excites into the Mg ground state by emitting gamma-ray photons
with the characteristic energy of 1809 keV . \Al is produced
almost exclusively by proton capture on $^{25}$Mg in a
sufficiently hot environment. \Al origin is dominated by massive
star and core-collapse supernovae, and small part of \Al is
attributed to AGB stars and novae.

Therefore, $^{26}$Al and $^{60}$Fe would share at least some of
the same production sites, i.e. massive stars and supernovae. In
addition both are long-lived radioactive isotopes, so we believe
their gamma-ray distributions are similar as well. We derive the
ratio of \Fe/\Al, which can be directly compared with theoretical
predictions.

We also obtain the \Al spectrum in the Galaxy using three years of
INTEGRAL data which is shown in Figure 4. \Al flux is $ (2.99\pm
0.24)\times 10^{-4}\mathrm{ ph\ cm^{-2}\ s^{-1}\ rad^{-1}}$.
Combining the \Fe result in Figure 3, we find a flux ratio of
\Fe/\Al of $15\pm 5)\%$.

\begin{figure}
\centering
\includegraphics[width=9cm]{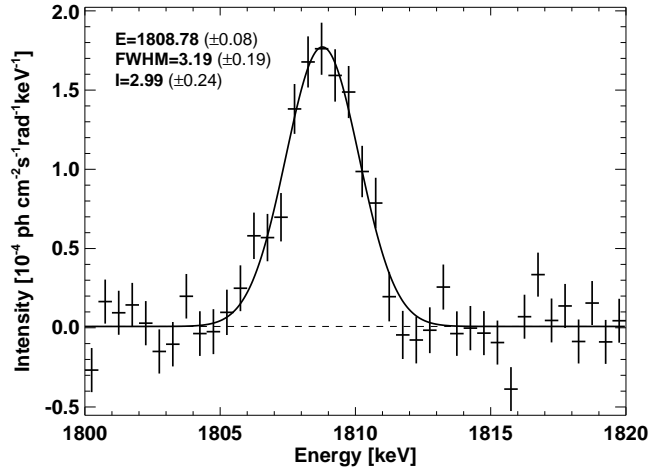}
\caption{\Al spectrum derived by INTEGRAL/SPI with 3 years of
data. \Al flux in the Galaxy is $ (2.99\pm 0.24)\times
10^{-4}\mathrm{ ph\ cm^{-2}\ s^{-1}\ rad^{-1}}$.  }
\end{figure}

\begin{table}\def~{\hphantom{0}}
  \begin{center}
  \caption{Different measurements of $^{60}$Fe/$^{26}$Al flux ratio}
  \label{tab:kd}
  \begin{tabular}{lcl}\hline
      Experiments &  $F(^{60}\mathrm{ Fe})/F(^{26}\mathrm{ Al})$ & references  \\
\hline
HEAO-3  & $0.09\pm 0.08$  & Mahoney et al. 1982 \\
SMM & $0.1\pm 0.08$    &  Leising \& Share 1994 \\
OSSE &$0.21 \pm 0.15$    &  Harris et al. 1997 \\
COMPTEL  &  $0.17\pm 0.135$     & Diehl et al. 1997  \\
GRIS &  $<0.14 (2\sigma)$  & Naya et al. 1998 \\
RHESSI  & $0.16\pm 0.13$ & Smith 2004 \\
SPI&  $0.15\pm 0.5$ & this work \\\hline
  \end{tabular}
 \end{center}
\end{table}

\begin{figure}
\centering
\includegraphics[width=12cm]{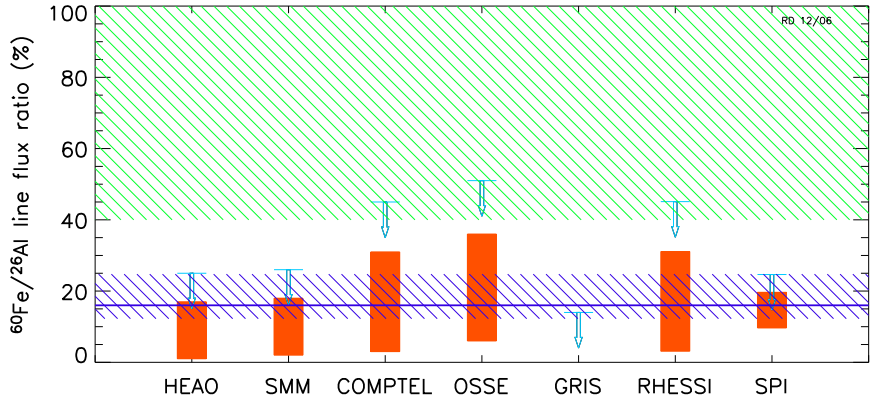}
\caption{Flux ratio of the gamma-ray lines from the two long-lived
radioactive isotopes $^{60}$Fe/$^{26}$Al from several
observations, including our SPI result (also see Table 1, from
Wang et al. 2007), with upper limits shown at 2$\sigma$ for all
reported values, and comparison with the recent theoretical
estimates (the upper hatched region from Prantzos 2004; the
straight line taken from Timmes et al. 1995; the lower hatched
region, see Limongi \& Chieffi 2006). Our present work finds the
line flux ratio to be $ (15\pm 5)\%$. See more details in the
text. }
\end{figure}

Many experiments and efforts were made to measure the
$^{60}$Fe/$^{26}$Al flux ratio, and we now provide the most
significant detection to date (see Table1 and Figure 5). In the
same time, different theoretical models have predicted the ratio
of $^{60}$Fe/$^{26}$Al. Timmes et al. (1995) published the first
detailed theoretical prediction. In their paper, they combined a
model for $^{26}$Al and $^{60}$Fe nucleosynthesis in supernova
explosions with a model of chemical evolution, giving a gamma-ray
flux ratio $F(^{60}{\rm Fe})/F(^{26}{\rm Al})= 0.16\pm 0.12$.
Since 2002, theoreticians have improved various aspects of the
stellar-evolution models, including improved stellar wind models
and the corresponding mass loss effects on stellar structure and
evolution, of mixing effects from rotation, and also updated
nuclear cross sections in the nucleosynthesis parts of the models.
As a result, predicted flux ratios $^{60}$Fe/ $^{26}$Al rather
fell into the range $ 0.8\pm 0.4$ (Prantzos 2004, based on, e.g.
Rauscher et al. 2002, Limongi \& Chieffi 2003) -- such high values
would be inconsistent with several observational limits and our
SPI result. Limongi \& Chieffi (2006) combined their individual
yields, using a standard stellar-mass distribution function, to
produce an estimate of the $^{60}$Fe/ $^{26}$Al gamma-ray flux
ratio expected from massive stars. Their calculations yield a
lower prediction for the $^{60}$Fe/ $^{26}$Al flux ratio of $
0.185\pm 0.0625$, which is again consistent with the observational
constraints.

\section{Summary and discussion}

Now, we have detected both 1173 keV and 1332 keV lines of \Fe in
the Galaxy ( near 5 $\sigma$ significance) with the 3 years of
SPI/INTEGRAL data, which is the best results on detections of \Fe
in the Galaxy, and confirms its existence. The average $^{60}$Fe
line flux from the inner Galaxy region is $ (4.4\pm 0.9)\times
10^{-5}{\rm ph\ cm^{-2}\ s^{-1}\ rad^{-1}}$. From the same
observations and analysis procedure applied to $^{26}$Al, we find
a flux ratio of $^{60}$Fe/ $^{26}$Al of $(15 \pm 5.0) \%$.

Though large error bars and uncertainties exist, the original and
the latest theoretical prediction of the flux ratio of $^{60}$Fe/
$^{26}$Al are consistent with our SPI result. But improvements are
needed both in observations and theories. For gamma-ray astronomy,
more precise measurements of gamma-ray lines in the Galaxy are
required, especially for the $^{60}$Fe signals, which may require
the more SPI data and the development of next-generation gamma-ray
spectrometers/telescopes. Stellar evolution models have potential
for improvements in processes related to the production of
$^{60}$Fe and $^{26}$Al, e.g. convective layers in the inner
stars, wind models for WR and O stars and the possible effects of
stellar rotation (Hirschi et al. 2004). The nuclear physics still
has serious uncertainties for the productions of $^{26}$Al and
$^{60}$Fe. For example, the cross section of $^{12}{\rm
C}(\alpha,\gamma)^{16}$O is uncertain, which affects the
prediction of both $^{26}$Al and $^{60}$Fe; the situation of
$^{60}$Fe is strongly influenced by the cross sections of neutron
capture and $\beta$-decay which are purely theoretical: no
experimental data exist for the $^{59}{\rm Fe}(n,\gamma)$ and
$^{60}{\rm Fe}(n,\gamma)$ rates. Therefore, a concerted effort
among stellar models, nucleosynthesis theory, and gamma-ray
observations is required for a more satisfactory assessment of
$^{60}$Fe synthesis in the Galaxy.

\begin{acknowledgments}
We would like to acknowledge R. Diehl and M. Harris for the
discussions.
\end{acknowledgments}

\end{document}